\documentclass[journal=jpclcd,manuscript=article]{achemso}

\usepackage[version=3]{mhchem} 
\usepackage{graphicx}
\usepackage{amsmath, mathtools}
\usepackage{array}

\author{Jan-Niklas Boyn}
\author{Jiaze Xie}
\author{John S. Anderson}
\author{David A. Mazziotti}
\email{damazz@uchicago.edu}
\affiliation[The University of Chicago]{The James Franck Institute and The Department of Chemistry, The University of Chicago, Chicago, Illinois 60637 USA}

\title[]{Entangled Electrons Drive a non Superexchange Mechanism in a Cobalt Quinoid Dimer Complex}

\begin{document}

\begin{tocentry}
\includegraphics[]{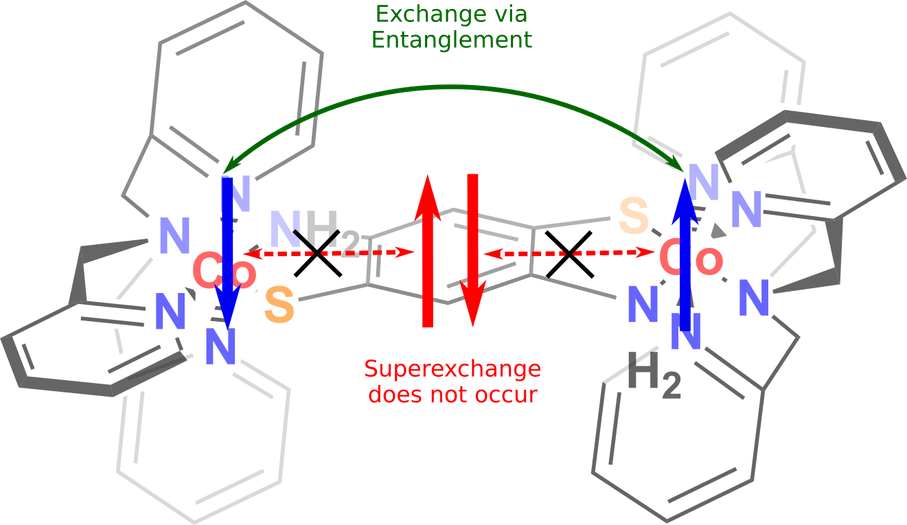}
\end{tocentry}

\begin{abstract}
A central theme in chemistry is the understanding of the mechanisms that drive chemical transformations.  A well-known, highly cited mechanism in organometallic chemistry is the superexchange mechanism in which unpaired electrons on two or more metal centers interact through an electron pair of the bridging ligand.  We use a combination of novel synthesis and computation to show that such interactions may in fact occur by a more direct mechanism than superexchange that is based on direct quantum entanglement of the two metal centers.  Specifically, we synthesize and experimentally characterize a novel cobalt dimer complex with benzoquinoid bridging ligands and investigate its electronic structure with the variational two-electron reduced density matrix method using large active spaces.  The result draws novel connections between inorganic mechanisms and quantum entanglement, thereby opening new possibilities for the design of strongly correlated organometallic compounds whose magnetic and spin properties have applications in superconductors, energy storage, thermoelectrics, and spintronics.
\end{abstract}

\textit{Introduction} - Transition metal dimer complexes have been an area of significant research interest in recent years owing to their wide range of tunability for different applications. Indeed, metal centers bridged by tunable organic ligands have become leading candidates in the development of single molecule magnets (SMM) and extended framework magnetic materials\cite{SCM,SCM2,QCompSMM,singlechain,2dsemiq,2d2,chainMn}. Ideal candidates for SMMs have large relaxation energy barriers leading to consequently large relaxation times\cite{Blagg2013,Co2Rev, OrganometallicsRev}. Magnetic coupling between metal centers in bridging coordination compounds is thought to be mediated though interactions with the electron density on the bridging ligand: a superexchange mechanism between two paramagnetic metal centers when the ligand is diamagnetic, or a direct exchange interaction when the ligand is paramagnetic\cite{Superexchange, Exchange, Exchange2, THarris1, THarris2, NatureVanadium}. Exchange interactions are a quantum mechanical phenomenon that arise from the entanglement of electrons across multiple centers. Formally, quantum entanglement  occurs in a pure state when the total wave function cannot be expressed as a product of the wave functions of the entangled centers\cite{entanglement1, entanglement2}. Superexchange interactions have been illustrated and studied in transition mental complexes with oxido linkers, where the magnitude of the interaction facilitated via the one atom bridge can be significant\cite{FirstSMM, SMM2, SMM3, SMM4, SMM5, SMM6, Halide}. Exchange interactions have important implications for the magnetic properties of electronic systems with positive exchange interactions favoring parallel spins and negative exchange energies promoting the pairing of spins.  \\

Since the discovery of the first SMM, Mn$_{12}$O$_{12}$(O$_2$CMe)$_{16}$(H$_2$O)$_4$, a Mn cluster compound\cite{FirstSMM}, researchers have been interested in the synthesis of transition metal dimers and clusters bridged by organic ligands with the aim of maximizing the ground-state multiplicity\cite{Rev1,Rev2}. Recent research has led to the successful synthesis and experimental characterization of a large range of compounds, covering a wide range of transition metal centers, as well as various lanthanide centers\cite{FeQD, LnRev, NatureVanadium, Dy, SMMHelicate}. Complexes with bis(bidentate) benzoquinoid based ligands with oxygen, sulfur and nitrogen donor atoms have shown particularly promising results\cite{THarris2, Rev2, similar, similar1, similar2, similar5, similar6, similar7, similar8, similar9, similar10, similar11, THarris3}. The focus of this synthetic work has been the tuning of magnetic exchange interaction in these systems, aiming to yield high spin ground states and large exchange constants J through changes of orbital structure and electron density on the bridging ligand, and more recently by using radical ligands to enable direct exchange coupling\cite{Rev1,THarris1,similar3,similar4,THarris4,THarris5}. However, theoretical and modelling work aimed at elucidating the electronic structure and nature of exchange coupling present in these compounds has been sparse with the little published work mostly relying on density functional theory (DFT) calculations\cite{ExpCAS, ExpDFT, Halide, ExpDFT2, ExpDFT3, ExpDFT4, ExpDFT5}, which are well known to struggle with strongly correlated systems\cite{DFTMed, DFTfail, DFTfail2, DFTfail3, DFTfail4}. \\

In this letter we present the synthesis, experimental characterization and computational electronic structure investigation of the novel dimeric complex [(CoTPA)$_2$DADT][BF$_4$]$_2$ (TPA: tris(2-pyridylmethyl)amine, DADT: 2,5-diaminobenzene-1,4-bis(thiolate)). The electronic structure is investigated using the active-space variational 2-RDM method (V2RDM)\cite{RDM2, V2RDMrev, V2RDM, V2RDM2, V2RDM3, V2RDM4, V2RDMOS, V2RDM5, V2RDM6, V2RDM7, V2RDM8, V2RDM9, V2exp, V2exp4, V2New}. We elucidate the nature of the bonding and exchange in [(CoTPA)$_2$DADT]$^{2+}$ and provide evidence that, surprisingly, superexchange is not responsible for the magnetic coupling between the two cobalt centers and that exchange is not mediated by the orbitals of the diamagnetic benzoquinoid ligand. \\

\begin{scheme}
  \includegraphics[scale=0.5]{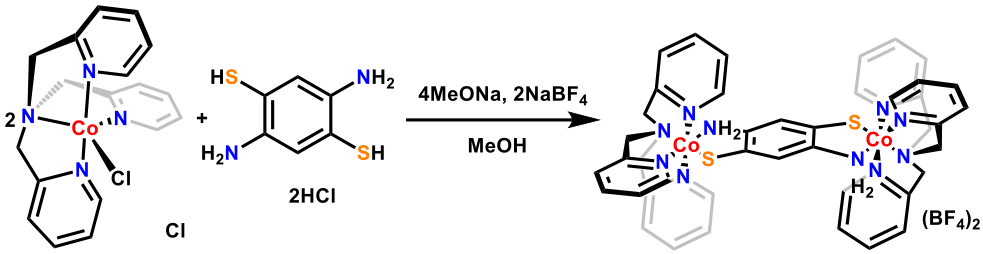}
  \caption{Synthesis scheme for [(CoTPA)$_2$DADT][BF$_4$]$_2$.}
  \label{sch:Synthesis}
\end{scheme}

\begin{figure}
    \centering
    \includegraphics[scale=0.6]{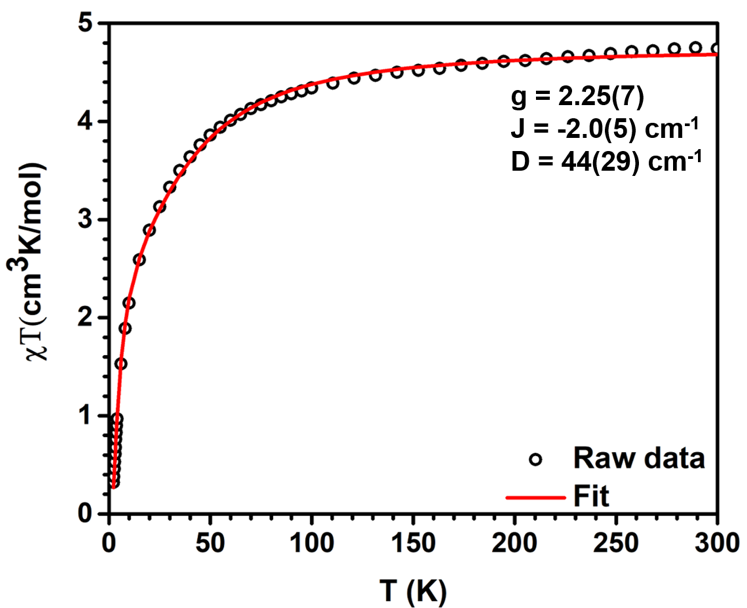}
    \caption{Varied-temperature $\chi T$ data for [(CoTPA)$_2$DADT][BF$_4$]$_2$, collected under an applied field of 1000 Oe from 300 to 1.8 K. Red line is the fit as described in the text.}
    \label{fig:SQUID}
\end{figure}

\textit{Results} - [(CoTPA)$_2$DADT][BF$_4$]$_2$ was synthesized by mixing (TPA)CoCl$_2$ with the hydrochloride salt of DADT and base (Scheme \ref{sch:Synthesis}) and was characterized with a suite of methods including single crystal X-ray diffraction (SXRD) and magnetic measurements. Variable temperature magnetic susceptibility superconducting quantum interference device (SQUID) measurements were carried out to probe the magnetic interactions in [(CoTPA)$_2$DADT][BF$_4$]$_2$ and determine the electronic ground state. The resulting magnetic susceptibility is shown in the plot of $\chi T$ vs $T$ in Fig. \ref{fig:SQUID}. We observe a singlet electronic state for the [(CoTPA)$_2$DADT]$^{2+}$ complex in the solid state low T limit. Due to possible inter-molecular interactions and zero-field splitting effects, this likely may not reflect the electronic ground state of the molecular species. The decrease in $\chi T$ from 300K to 100K arises from weak antiferromagnetic exchange coupling between the two high-spin Co centers in the molecular species. Upon initial inspection of the data, we assigned the source of this antiferromagnetic coupling as superexchange interactions mediated by the DADT linker, in accordance with similar assignments in the literature for various transition metal dimers and SMM candidates\cite{Halide, NatureVanadium, THarris1, similar1, similar9, THarris4, similar7, similar6, similar3}. Unlike in a direct exchange mechanism between nearest neighbor atoms, in the superexchange mechanism a diamagnetic ligand facilitates antiferromagnetic coupling between the two open-shell transition metal cations. Spins on each metal center are effectively paired through interaction with an electron pair localized on the bridging ligand, enabling exchange coupling over longer distances. This mechanism is schematically illustrated in Scheme \ref{sch:superexchange}. To quantify this antiferromagnetic interaction, we fit the data to the Van Vleck equation according to the spin Hamiltonian $\hat{H} = -2\text{J}(\hat{S}_{Co_1} \cdot \hat{S}_{Co_2})$, giving an exchange constant of $\text{J} = -2.0 (5)$ cm$^{-1}$\cite{VanVleck}. \\

\begin{scheme}
    \begin{minipage}[b]{0.45\textwidth}
        \centering
        \includegraphics[scale=0.48]{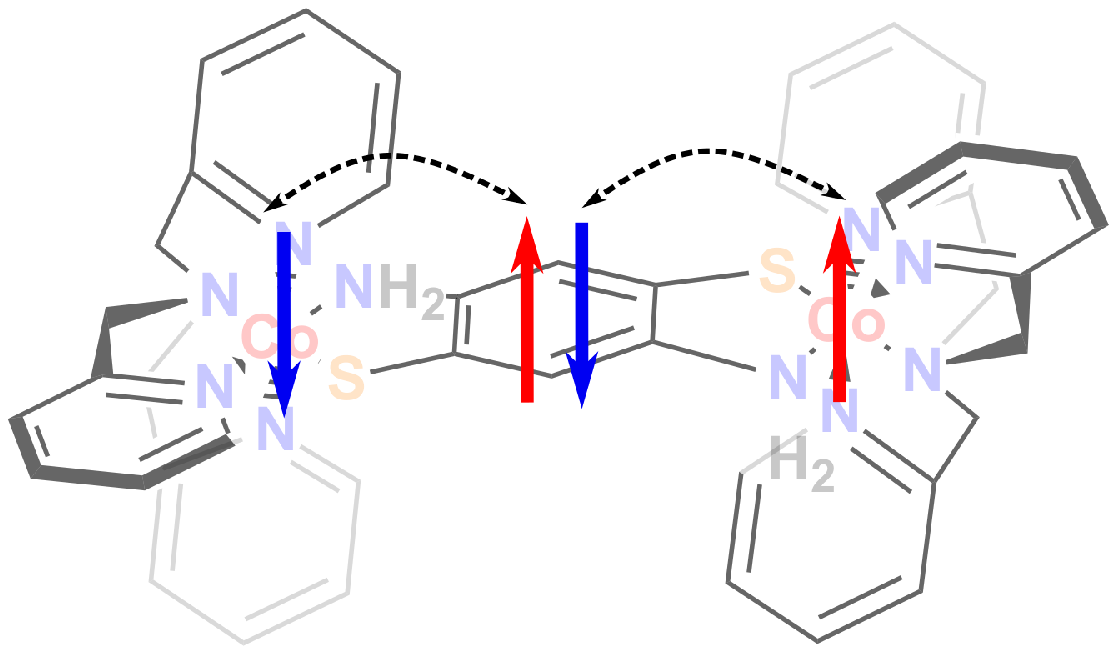}
    \end{minipage}
    \begin{minipage}[b]{0.45\textwidth}
        \centering
        \includegraphics[scale=0.35]{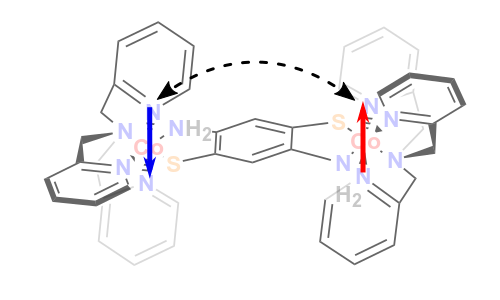}
    \end{minipage}
    \caption{Schematic representation of superexchange (left) and EPR-like direct exchange arising from the entanglement of the electrons across the two Co centers (right) in the [(CoTPA)$_2$DADT]$^{2+}$ complex.}
    \label{sch:superexchange}
\end{scheme}

For the electronic structure calculations the complex geometry is obtained from the SXRD structure and relaxed with DFT using the B3LYP functional in a 6-31G* basis set. Electronic energies and natural orbital occupation numbers (NON) were obtained for the singlet state with V2RDM complete active space self-consistent field (CASSCF) calculations in the Maple Quantum Chemistry Package\cite{maple,QCP} (QCP) for [12,10], [14,14], [20,20] and [24,24] active spaces and 6-31G basis sets\cite{basis}. The data is presented in the supporting information. A [14,14] active space can be inferred to be sufficient to describe the correlation and electronic structure of this system, and all further V2RDM calculations are performed using this active space. \\

V2RDM calculations in Table \ref{tab:V2RDMSpin} reveal the spin state splitting and determine the electronic ground state of the isolated molecule. The calculations elucidate the strongly correlated open-shell nature of this system showing significant partial occupations in the frontier NOs and a triplet ground state with a low lying quintet state separated by $\Delta E_{Q-T} = 7.77$ cm$^{-1}$ with inaccessible high energy singlet and septet states. Assuming the Heisenberg-Dirac-van-Vleck Hamiltonian for localized, weakly interacting electrons can be applied, the triplet-quintet gap corresponds to a J value of $\Delta E_{T-Q} = 4J$ or $J = -1.94$ cm$^{-1}$. This strongly suggests that the isolated molecule is in a triplet ground state and the experimentally measured antiferromagnetic exchange coupling between the two Co centers can be related to the gap between the triplet ground state and the low lying quintet state. Broken symmetry, unrestricted DFT calculations cannot account for the strong correlation in this system and fail to accurately predict its electronic structure with different functionals giving wildly varying results (refer to the supporting information). \\

\begin{table}
    \centering
    \begin{tabular}{c|cccc|ccc}
         & \multicolumn{4}{c|}{Dimer} & \multicolumn{3}{c}{Capped}\\
         & S & T & Q & ST & S & T & Q \\
         \hline
         E$_{rel}$(cm$^{-1}$) & 8970 & 0 & 7.77 & 25500 & 7790 & 0 & 196 \\
         $\lambda_{220}$ $|$ $\lambda_{203}$ & 1.99 & 1.99 & 1.99 & 1.82 & 1.98 & 1.99 & 1.99 \\
         $\lambda_{221}$ $|$ $\lambda_{204}$ & 1.84 & 1.68 & 1.67 & 1.75 & 1.88 & 1.68 & 1.65 \\
         $\lambda_{222}$ $|$ $\lambda_{205}$ & 1.79 & 1.58 & 1.58 & 1.75 & 1.85 & 1.63 & 1.61 \\
         $\lambda_{223}$ $|$ $\lambda_{206}$ & 1.78 & 1.50 & 1.50 & 1.00 & 1.85 & 1.52 & 1.47 \\
         $\lambda_{224}$ $|$ $\lambda_{207}$ & 1.00 & 1.12 & 1.12 & 1.00 & 1.00 & 1.08 & 1.13 \\
         $\lambda_{225}$ $|$ $\lambda_{208}$ & 1.00 & 1.10 & 1.11 & 1.00 & 0.94 & 1.06 & 1.12 \\
         $\lambda_{226}$ $|$ $\lambda_{209}$ & 0.56 & 1.00 & 1.00 & 1.00 & 0.45 & 1.00 & 1.00 \\
         $\lambda_{227}$ $|$ $\lambda_{210}$ & 0.02 & 0.01 & 0.01 & 0.26 & 0.02 & 0.01 & 0.01 \\
         $\lambda_{228}$ $|$ $\lambda_{211}$ & 0.01 & 0.01 & 0.01 & 0.25 & 0.02 & 0.01 & 0.01 \\
    \end{tabular}
    \caption{Energies relative to the triplet ground state, and NON for the dimeric complex and the capped Co centers. [14,14] V2RDM CASSCF calculations carried out in Maple QCP with a 6-31G basis set.}
    \label{tab:V2RDMSpin}
\end{table}

The nature of the spin state splittings can be rationalized with the help of the NON of the partially occupied frontier NOs and by inspection of the electron density. While the triplet and quintet state exhibit identical NON, singlet and septet states show significant redistribution of electrons within the frontier NOs. In the case of the singlet state we observe an open-shell bi-radical state and a notable decrease in partial occupancy, suggesting that the destabilization of the singlet by $\Delta E_{S-T} = 9320$ cm$^{-1}$ compared to the triplet ground state arises from a loss of strong correlation. The septet state ($\Delta E_{ST-T} = 25500$ cm$^{-1}$) requires the unfavorable flipping of an additional spin and population of higher energy orbitals.
Knowing the nature of the symmetry breaking in the frontier NOs, we perform DFT calculations at the B3LYP/6-31G* level of theory using a fragment guess approach, yielding $\Delta E_{Q-T} = -2.11$ cm$^{-1}$. While this splitting is of the correct order of magnitude, its sign is opposite to that of experimental and V2RDM results, predicting ferromagnetic exchange coupling instead of the observed antiferromagnetic coupling. \\

\begin{figure}
    \centering
    \includegraphics[scale=0.3]{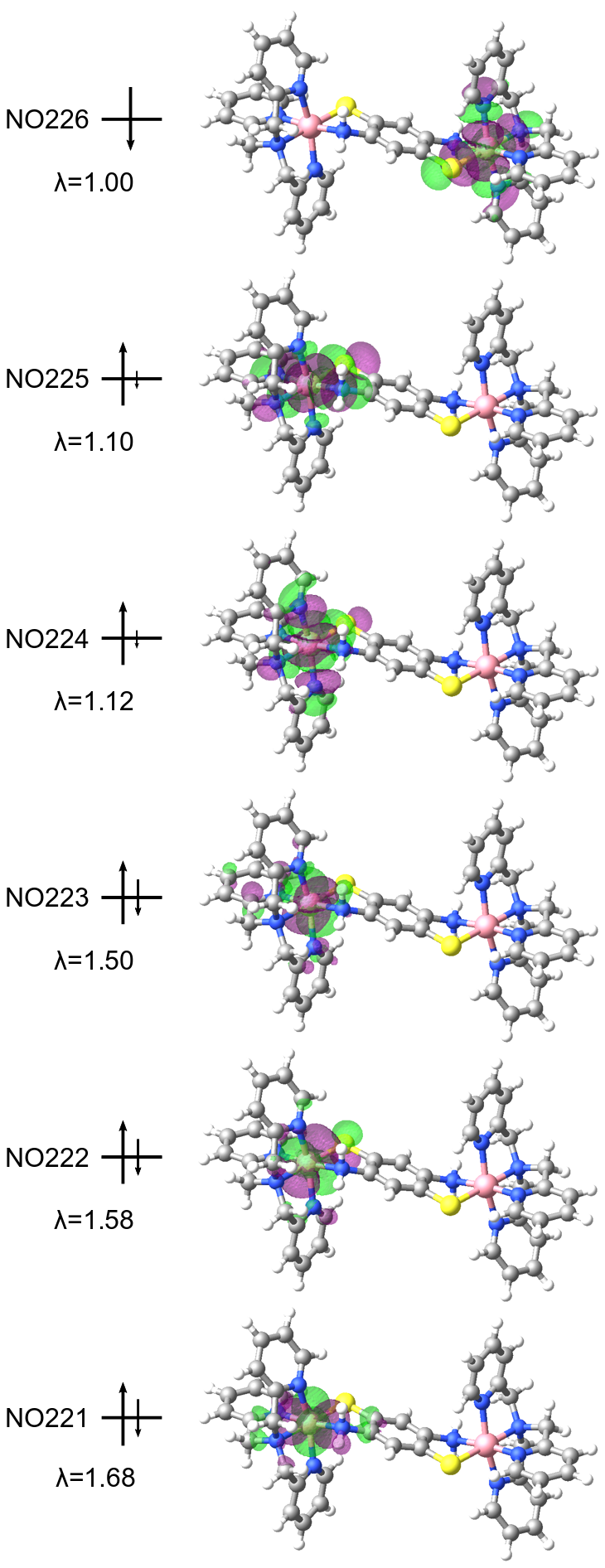}
    \caption{NO diagram for the triplet ground state, displaying the electron density in the partially occupied frontier NOs and their NON.}
    \label{fig:natorb_diag}
\end{figure}

We can elucidate the nature of the exchange coupling examining the frontier NO densities. V2RDM CASSCF minimization leads to a result of severely broken symmetry with all frontier NOs except one localized on one Co center. In the triplet and quintet case, shown in Figure \ref{fig:natorb_diag}, this yields a strongly correlated system of one Co center with net spin of $S=3/2$ and one center with $S=1/2$. In a superexchange picture the spins on the Co centers would couple by interacting with the electron density on the linker. However, the electron densities in all frontier NOs are highly localized on the two Co centers without contribution from the atomic orbitals of the bridging ligand. Hence, there cannot be an interaction of the unpaired electrons on the Co centers via the bridging ligand's benzene core and antiferromagnetic coupling is not facilitated by a superexchange mechanism. Rather, these results point to an Einstein-Podolsky-Rosen (EPR) like direct exchange mechanism between the unpaired electrons entangled at a distance across the Co centers. \\

\begin{figure}
    \centering
    \includegraphics[scale=0.3]{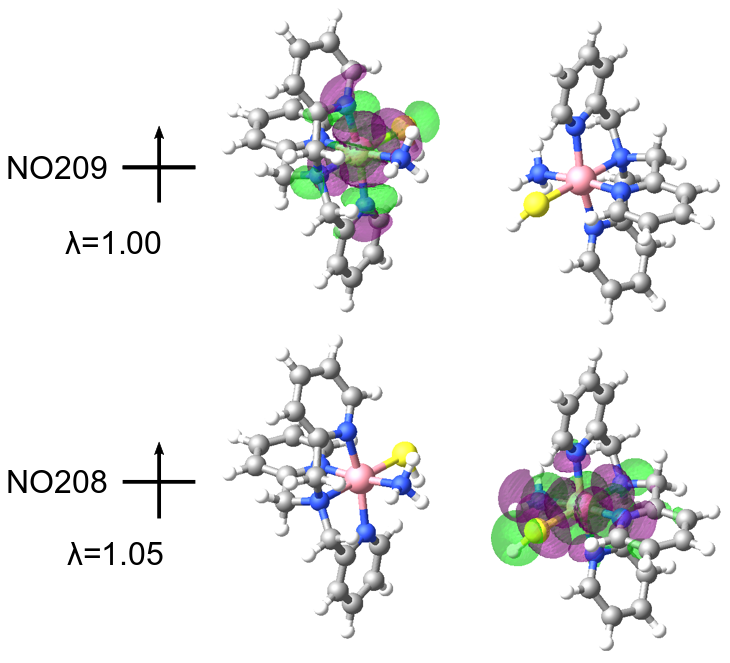}
    \caption{Diagram of NOs 208 and 209 for the two Co centers without linker. The frontier NOs remain unchanged by removal of the bridging ligand, suggest a EPR-like direct exchange mechanism, rather than superexchange facilitated by the linker.}
    \label{fig:caporbs}
\end{figure}

This conclusion is supported by calculations performed on the two Co centers fixed at the same Co-Co distances as those in the complex but with the bridging ligand removed and the nitrogen and sulfur atoms coordinated to the Co centers capped with hydrogens (referred to as capped structure). The resulting electronic structure displays similar NON (right hand column of Table \ref{tab:V2RDMSpin}) and the atomic orbitals involved in the frontier NOs remain unchanged upon removal of the linker (Figure \ref{fig:caporbs}). Calculations on the spin manifold show a triplet ground state with a low lying quintet state, separated by $\Delta E_{Q-T} = 196$ cm$^{-1}$, and again a higher lying, inaccessible singlet at $\Delta E_{T-S} = 7790$ cm$^{-1}$. These splittings are very close to the ones obtained for the dimer, in fact removal of the linker increases the exchange coupling constant from $J = -1.94$ cm$^{-1}$ to $J = -48.9$ cm$^{-1}$. This suggests, that not only is the linker not involved in the exchange mechanism via a superexchange pathway but rather its electron density plays a minimally shielding role, slightly reducing the exchange coupling compared to two Co centers directly entangled at a distance. \\

While [(CoTPA)$_2$DADT][BF$_4$]$_2$ exhibits weak exchange coupling, much stronger couplings have been observed in radical bridged complexes\cite{Rev1,THarris1,similar3,similar4,THarris4,THarris5}. To investigate the change to the electronic structure and magnetic exchange mechanism in the case of a radical bridging ligand additional calculations are performed on the vertically ionized species (center columns of Table \ref{tab:vertdata}). Mulliken charges on the Co centers remain unchanged and as expected ionization occurs from the DADT ligand. In the NO picture this corresponds to the removal of an electron from the inner NOs, yielding 5 approximately singly occupied NOs in all spin states.  \\

\begin{figure}
    \centering
    \includegraphics[scale=0.25]{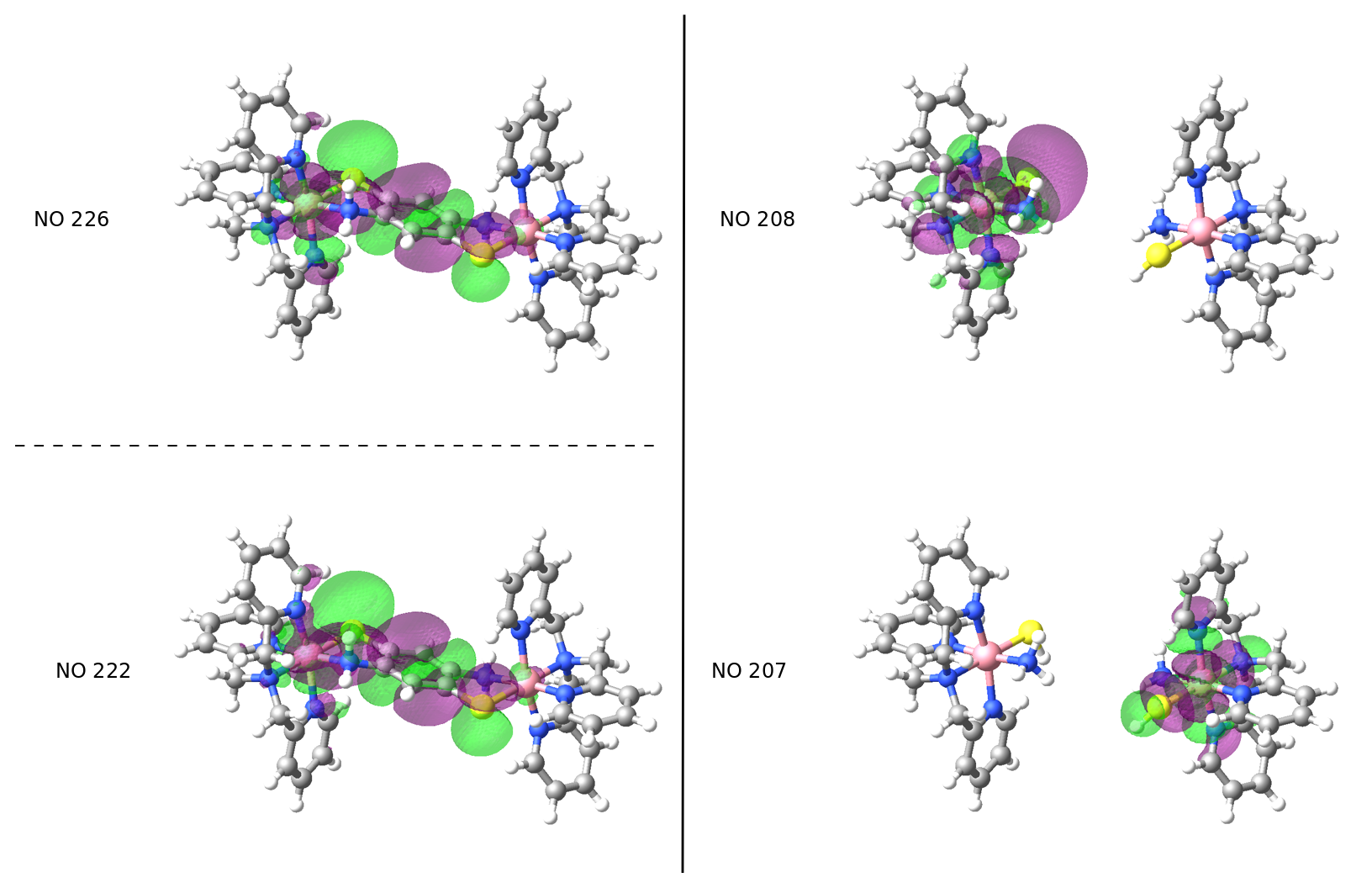}
    \caption{Left hand side: NOs 222 and 226 after vertical ionization to the +3 state. In contrast to the +2 complex, localization of the frontier NOs on the Co centers is lost, and the singly occupied frontier NOs show delocalization across the bridging ligand. Right hand side: Frontier NOs 207 and 208 for the capped arrangement after vertical ionization. The frontier NOs resemble those of the +2 state.}
    \label{fig:orbvert}
\end{figure}

The presence of the radical linker substantially changes the frontier orbital picture. While the majority of the NOs remain unchanged from the 2+ state and the system retains broken symmetry, NOs 222 and 226 show significant delocalization of the electron density across the linker, coupling the two cobalt centers (shown for the doublet on the left hand side of Figure \ref{fig:orbvert}). It becomes clear that in the case of the radical benzoquinoid bridging ligand, the two Co centers are no longer isolated and exchange may be mediated by the electron density on the bridge. V2RDM calculations show a doublet ground state and low lying, accessible quartet and sextet states, separated by $\Delta E_{Q-D} = 12.6$ cm$^{-1}$ and $\Delta E_{S-D} = 223$ cm$^{-1}$. The octet state is inaccessible ($\Delta E_{O-D} = 20800$ cm$^{-1}$). In agreement with the experimental observations this shows that involvement of a radical bridging ligand does indeed lead to spin state splittings and consequently exchange couplings of larger magnitude compared to the diamagnetic case, and this stronger interaction is mediated by the atomic orbitals of the linker. This image of radical linker orbital involvement is reinforced by calculations performed on a capped version of the vertically ionized complex (right columns of Table \ref{tab:vertdata}). In this case the electronic structure differs significantly from the dimer and the NOs resemble those of the 2+ case (the right hand side of Figure \ref{fig:orbvert}). These results help rationalize recent findings showing radical benzoquinoid ligands to yield significant higher $J$ compared to their neutral counterparts. \\

 \begin{table}
    \centering
    \begin{tabular}{c|cccc|ccc}
         & \multicolumn{4}{c|}{Dimer} & \multicolumn{3}{c}{Capped}\\
         & D & Q & S & O & D & Q & S \\
         \hline
         E$_{rel}$(cm$^{-1}$) & 0 & 12.6 & 223 & 25400 & 0 & 3270 & 20800 \\
         $\lambda_{220}$ $|$ $\lambda_{203}$  & 1.95 & 1.95 & 1.95 & 1.73 & 1.95 & 1.94 & 1.84 \\
         $\lambda_{221}$ $|$ $\lambda_{204}$  & 1.95 & 1.95 & 1.95 & 1.73 & 1.95 & 1.94 & 1.82  \\
         $\lambda_{222}$ $|$ $\lambda_{205}$  & 1.15 & 1.14 & 1.02 & 1.00 & 1.80 & 1.77 & 1.49  \\
         $\lambda_{223}$ $|$ $\lambda_{206}$  & 1.00 & 1.01 & 1.00 & 1.00 & 1.71 & 1.47 & 1.22  \\
         $\lambda_{224}$ $|$ $\lambda_{207}$  & 1.00 & 1.00 & 1.00 & 1.00 & 0.80 & 0.99 & 1.16  \\
         $\lambda_{225}$ $|$ $\lambda_{208}$  & 0.98 & 0.98 & 0.99 & 1.00 & 0.69 & 0.77 & 1.00  \\
         $\lambda_{226}$ $|$ $\lambda_{209}$  & 0.87 & 0.88 & 0.99 & 1.00 & 0.05 & 0.06 & 0.16  \\
         $\lambda_{227}$ $|$ $\lambda_{210}$  & 0.05 & 0.05 & 0.05 & 0.27 & 0.05 & 0.05 & 0.16  \\
         $\lambda_{228}$ $|$ $\lambda_{211}$  & 0.05 & 0.05 & 0.05 & 0.27 & 0.04 & 0.05 & 0.15  \\
    \end{tabular}
    \caption{Energies relative to the doublet ground state, and NON for the vertically ionized 3+ dimeric complex and the capped Co centers fixed at the dimer geometry but without the bridging ligand. [13,14] V2RDM CASSCF calculations carried out in Maple QCP with a 6-31G basis set.}
    \label{tab:vertdata}
\end{table}

\textit{Conclusions} -
Through the use of large active space CASSCF calculations we render a quantum mechanical picture of the exchange interaction in a novel benzoquinoid bridged cobalt dimer complex, [(CoTPA)$_2$DADT]$^{2+}$, allowing us to attribute the experimentally characterized exchange interaction to entanglement rather the commonly accepted superexchange picture. V2RDM calculations correctly reproduce the experimental exchange constant $J$, showing a strongly symmetry broken wave-function with highly localized singly occupied frontier NOs centered on the Co atoms with no involvement of the benzoquinoid linker, firmly suggesting that superexchange, the commonly accepted mechanism for long distance spin pairing in bridged transition metal and lanthanide complexes and clusters, is not responsible for the antiferromagnetic interaction in this system. Instead, our results suggest an EPR-like, direct exchange mechanism is operating over long distances between the two Co centers in [(CoTPA)$_2$DADT]$^{2+}$, illustrated in Scheme \ref{sch:superexchange}. This is further corroborated by V2RDM calculations on the capped Co centers fixed to their positions in the [(CoTPA)$_2$DADT]$^{2+}$ complex, revealing no major changes in the electronic structure upon removal of DADT. and instead leading to an increase in the magnitude of the exchange coupling, suggesting the presence of the linker may actually shield the exchange interaction between the two metal centers. This is in stark contrast to the electronic structure following vertical ionization, removing an electron from DADT and yielding a radical linker, enabling its participation in the exchange pathway. This is reflected in the frontier NO densities, which are delocalized over both Co centers as well as the linker allowing exchange coupling between the metal centers via the bridge, in line with assignments in the literature. \\
The work presented here offers compelling evidence that superexchange interactions over long ligand distances may not be a dominant factor in exchange coupling and instead correlation and entanglement may be the chief contributors. The results do, however, confirm a radical linker does, indeed, mediate the exchange interaction between two paramagnetic metal centers. Lastly, this work emphasizes the importance of large active space CASSCF calculations which allow for symmetry breaking in the wave function and account for strong correlation in the design of novel complexes with desirable electronic properties, with V2RDM results differing significantly from predictions obtained by traditional DFT functionals. The insight gained from these calculations can be especially useful in cases where the interaction between valence electrons is assumed to play a key role. \\

\begin{acknowledgement}
J. S. A. and D. A. M. gratefully acknowledge support for this work from the U.S. Department of Energy, Office of Science, Office of Basic Energy Sciences, under Award No. DE-SC0019215. D. A. M. gratefully acknowledges the National Science Foundation (NSF) Grant No. CHE-1152425, and the United States Army Research Office (ARO) Grant No. W911NF-16-1-0152. J. S. A. also acknowledges support from 3M through a NTFA grant.
\end{acknowledgement}

\begin{suppinfo}

Experimental procedures and characterization data, as well as computational methodology and data can be found in the SI.

\end{suppinfo}

\bibliography{CoDimer_JPCL}

\end{document}